\documentclass[fleqn,twoside]{article}
\usepackage{espcrc2}
\usepackage{epsfig}

\usepackage{graphicx}
\usepackage[figuresright]{rotating}

\newcommand{\AmS}{{\protect\the\textfont2
  A\kern-.1667em\lower.5ex\hbox{M}\kern-.125emS}}

\title{The apeNEXT Project}
\author{F. Bodin\address[irisa]{IRISA/INRIA, Campus Univ. de Beaulieu, Rennes},
	P. Boucaud\address[orsay]{LPT, University of Paris Sud, Orsay},
	N. Cabibbo\address[roma1]{INFN, Sezione di Roma},
	F. Calvayrac\address[mans]{LPEC, Univ. du Maine, Le Mans},
  	M. Della Morte\address[milano]{Physics Dept., University of Milano
        Bicocca and INFN, sezione di Milano, Italy}$^{,}$\address[desy]{DESY
        Zeuthen, Germany}, 
	R. De Pietri\address[parma]{Physics Dept., University of Parma and
        INFN, gruppo collegato di Parma}
	P. De Riso\address[roma2]{Physics Dept., University of Roma
	``Tor Vergata'' and INFN, Sezione di Roma II},
	F. Di Carlo\addressmark[roma1],
	F. Di Renzo\addressmark[parma],
	W. Errico\address[pisa]{INFN, Sezione di Pisa},
	R. Frezzotti\addressmark[milano],
	U. Gensch\addressmark[desy],
	T. Giorgino\addressmark[desy],
	M. Guagnelli\addressmark[roma2],
	N. Herv\'e\addressmark[irisa],
	H. Kaldass\addressmark[desy]
	\thanks{on leave of absence from University of Texas, Austin},
	A. Lonardo\addressmark[roma1],	
	M. Lukyanov\addressmark[desy]
	\thanks{on leave of absence from JINR, Dubna},	
	G. Magazz\'u\addressmark[pisa],
	J. Micheli\addressmark[orsay],
	V. Mor\'enas\address[clermont]{LPC, Univ. Blaise Pascal and
	IN2P3, Clermont}
	L. Mori\addressmark[parma],
	F. Palombi\addressmark[roma2],
	N. Paschedag\addressmark[desy],
	O. P\'ene\addressmark[orsay],
	R. Petronzio\addressmark[roma2],
	D. Pleiter\address[nic]{NIC/DESY Zeuthen},
	F. Rapuano\addressmark[roma1],
	D. Rossetti\addressmark[roma1],
	L. Sartori\addressmark[pisa],
	H. Simma\addressmark[desy],
	F. Schifano\addressmark[pisa],
	R. Tripiccione\address[ferrara]{Physics Dept., University of Ferrara
	and INFN, Sezione di Ferrara},
	P. Vicini\addressmark[roma1]}
                                            
\begin{document}

\begin{abstract}
apeNEXT is a new generation APE processor, optimized for LGT simulations.
The new project follows the basic ideas of previous APE machines and develops
simple and cheap parallel systems with multi-Tflops processing power.
This paper describes the main features of this new development.
\vspace{1pc}
\end{abstract}

\maketitle
\section{OVERVIEW}
Computer power requirements for lattice gauge theory (LGT)
simulations  have increased exponentially
in the last 20 years. Estimates of these requirements, in terms of computing
power and memory needs for large scale
simulations, such as studies of the hadronic spectrum with dynamical
fermions close
to the physical limit have been published
recently \cite{ecfa} and extensively debated at this conference
\cite{panel}. It is increasingly clear that unrestricted access to
computing resources of several sustained Tflops will be needed in
the next few years. For about 15 years, a large
fraction of the compute cycles used by LGT simulations have been provided
by LGT-optimized massively parallel processors.
However, presently available systems (see
\cite{norman1999} for a recent review) are  unable to reach the
required performance target. The development of
yet another generation of dedicated LGT engines is still one open option,
even if different approaches to the problem are currently
discussed \cite{martino}.

This paper describes in detail one such project, apeNEXT, carried out jointly
by DESY, INFN and the University of Paris-Sud (a similar project is described
in \cite{norman2001}). This paper is structured as follows: the next section
introduces requirements and guidelines of the apeNEXT architecture. Section 3
outlines the proposed structure of the new system, at the hardware level, while
section 4 discusses software issues. Section 5 is a brief  status report.

\section{PROJECT REQUIREMENTS AND GUIDELINES} Our main goal is the development
of a massively parallel system for ``compute intensive'' and ``memory
intensive'' LGT simulations. The former case refers to  dynamical fermions on
not too large lattices (typical sizes of $48^3 \times 96$ corresponding to
system of $L = 2 \ldots 4 \: fm$ and  $a = 0.1 \ldots 0.05 \: fm$) and small
dynamic quark masses, while the latter case corresponds to quenched simulations
on very large lattices ($100^3 \times 100 \div 200$) and large $\beta$ ($L =
1.5 \ldots 2.0 \: fm$ and $a = 0.1 \ldots 0.02 \: fm$) relevant for $b$ physics
with limited extrapolation in the mass of the heavy quark.

These physical requirements translate into the capability to handle online data
structures of the order of several hundred Gbyte to 1 Tbyte, allowing the use
of optimized algorithms that trade memory for performance, and to transfer from
disk to memory and viceversa the tens of Tbyte corresponding to the full set of
propagators needed for weak interaction studies (see \cite{propo} for an
accurate estimate of these figures). In terms of computing power, we
aim for systems delivering several TFlops. 

We try to minimize architectural changes with respect
to APEmille and to leverage on technology improvements, but
we introduce improvements in some key architectural areas:
\begin{itemize}\itemsep -1mm
\item Floating point arithmetics is upgraded to double precision.
\item Typical LGT kernels have a ratio $R \simeq 4 \cdots 6$
between the number of floating point operations and the required
operands. APEmille was able to provide enough memory bandwidth
without any intermediate storage hierarchy between registers and main memory.
In apeNEXT longer memory latencies (in units of processor
clock cycles) reduce effective bandwidths for data bursts of the
size of gauge matrices or spin-color fields.
We introduce additional features to handle this problem.
\item APEmille is not able to overlap remote data transfer with computation, so
the system must idle waiting for data from remote nodes to arrive. This
loss of performance can be partially avoided,
by allowing concurrent link activity and computations. We also 
support concurrent data transfers on more than one data link.
\end{itemize}

\section{THE apeNEXT ARCHITECTURE}

\begin{figure}[tb]
\centering
\vspace{0.0cm}
\epsfig{file=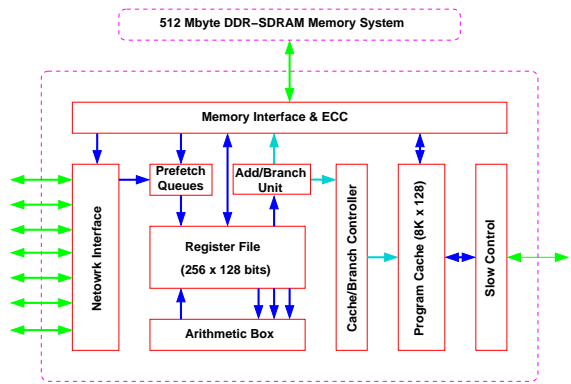, width=7.5cm}
\vspace{0.0cm}
\caption{Block diagram of J\&T, the single-chip processor used in apeNEXT.}
\label{fig:chip}
\end{figure}

The apeNEXT architecture is based on a three-dimensional grid of processing
nodes, connected by data links between first-neigbours with periodic boundary
conditions. Each node is a completely independent unit, with a processor and its
memory bank. It executes independently its own copy of the same program, so the
whole system is a SPMD processor.  Data transfers
between nodes follow the message-passing paradigm: the program on the
destination node must explicitely receive data sent by the source node. This is
done however with small latency ($\simeq 100 ns$).

apeNEXT is based on one building block, namely the J\&T processor. It
controls program flow, performs arithmetic operations and handles data
links to neighbour nodes. A block diagram is shown in figure \ref{fig:chip}.
J\&T has several similarities with previous APE processors:
\begin{itemize}\itemsep -1mm
\item An interface to main memory, through a data bus of 128 bits, plus error
detection and correction bits. The memory interface supports Double Data Rate
(DDR) Synchronous Dynamic Rams (SDRAM), that will be widely used in standard
PCs in the next two years. A set of 8 plus 1 memory chips provides 256 Mbytes
of main (program and data) memory for each node.
\item a large Register File  (RF), holding 256 registers of 64+64 bits. There
are three read ports, one write port and one bi-directional ports. Three read
ports feed data to the arithmetic box, results enter via the write port, and
the bi-dirirectional port exchanges data with memory and the queue system (see later). 
\item The arithmetic box performs the ``normal'' floating-point operation $a
\times b + c$ on complex (or pair of real) numbers, corresponding to 8 (or 4)
standard floating point operations per clock cycle. We obtain 1.6 Gflops peak
performance with a 200 MHz clock. Pipeline length is 10 clock cycles. The
arithmetic box also performs arithmetic and logic operations on pairs of 64 bit
integer values and conversions from/to floating point format.
\end{itemize}

The new features of J\&T include:
\begin{itemize}\itemsep -1mm
\item A software-controlled program cache, where heavily used compute
      kernels are pre-loaded.
\item An address-generation unit, which operates independently of the
      arithmetic unit.
\item A serial control interface for initialization, exception handling
      and debugging.
\item A queue system for pre-fetching local and remote data.
\item Seven data links with a bandwidth of 200 Mbytes/s each.
\end{itemize}
Here we discuss in some details
the memory-queue system and the data links.

The apeNEXT memory interface has a peak bandwidth of 3.2 Gbyte/sec
(one complex data word per clock cycle, able to sustain
algorithms with $R \le 4$). In practice,
a startup latency of about 12 cycles affects sustained
performances. We plan to solve this problem through a
program controlled prefetch mechanisms. During execution
of iteration $i$ of a critical kernel, data structures needed for iteration
$(i+1)$ are moved from memory to an intermediate storage element, a data queue,
close (in terms of bandwidth and latency) to the register file. The size of the
queue is 1024 complex words. The user program
starts data transfers from memory to the queue. Memory accesses are
local (from local memory to local queue)
or remote (from memory to the queue in a neighbour processor)
and hence
may have different transfer times. The queue system, however, ensures
delivery of the data to the receiving register file in the same order in
which memory accesses were scheduled.

Prefetch reduces substantially remote bandwith requirements.
Consider the evaluation of the Wilson-Dirac
operator,  for a lattice point on the border between two processing nodes. A
total of 12 complex data words (fermionic degrees of freedom) plus possibly 9 complex
data words (gauge fields) must be moved on a data link in the time interval
used for the evaluation of the operator, corresponding to $\simeq 320$ complex
normal operations. If we (unrealistically) assume 100\% efficiency, we require
a transfer rate of about 1 byte per clock cycle, i.e. 200
Mbytes/sec. Sites at the
edges (corners) of the physical region mapped onto each processor need two
(three) data transfers. 
If the links in different
directions operate concurrently bandwidth requirements are not increased.

The apeNEXT links are designed for the above requirements. 
Each
link delivers a block of data words to the destination queues. The transfer,
proceeeds concurrently with local node operations. An interlock
mechanisms stallss the receiving node if the latter tries to pop data not
yet arrived at the destination queue.

Data communication provides a synchronization mechanism: the user program starts a
data transfer to a remote node over one link. At the same time it enables data
reception from another link. If the sender is ahead in time with respect
to the receiver, the former will be forced to wait till synchronization is
achieved. In most cases relevant for LGT simulations, data sends and receive are
in opposite directions, but more general communication patterns, such as
systolic paths, can be established.

Each node has 7 data links. Six links make up the regular three-dimensional network,
while the ``seventh'' link is available for I/O to the host system (see later,
for more details). Each link uses Cyclic Redundancy Check (CRC) to identify
errors. Each data block is divided in packets of 128 bits (plus 16
CRC bits). Corrupted packets are re-trasmitted. 

apeNEXT machines use Processing Boards (PB) as the basic hardware building block.
Each PB has 16 processing elements (an array of  $4 \times 2 \times 2$
nodes). A set of 16 PBs is assembled in a Crate, a system of 
$4 \times 8 \times 8$ nodes with communication
links on the backplane. Crates are connected by cable links, building
larger systems of $4n \times 8 \times 8$ nodes. Each node 
is also connected to the slow control interface.  
Each PB can  be connected to a host
computer across the ``seventh link'', which handles heavy data I/O. 
The number of hosts and of PBs using the
``seventh link'' is arbitrary. Therefore the I/O bandwidth of apeNEXT
can be scaled according to the specific applications.

\section{SOFTWARE}
apeNEXT programs are written in the TAO programming language, also used
in earlier APE machines. Our goal is almost complete compatibility, in
the sense that old APEmille program will only need to be recompiled. Optimal
efficiency on apeNEXT will of course involve some machine specific tuning of
the codes. In particular programs will have to handle the queues, which
are accessible at TAO level.

We are also developing a C compiler, based on open source compilers.
Our target is to treat TAO and C on almost equal footing, also from the point of
view of efficiency. 
From first experiences with a prototype version
we hope that C codes reach performances comparable with TAO.

At the backend level of the compilation chain, we plan to use the instruction
scheduling techniques used in APEmille to optimize pipeline usage.
We are also in the process of adopting further optimization tools at the assembly level.

\section{PROJECT STATUS}
All apeNEXT hardware building blocks are in an advanced stage of development:
the first PB prototype has just been delivered, while the J\&T processor will
start to
be fabricated at a silicon foundry later this year. We expect all prototype
buiding blocks to be available next spring. Our next important goal is a
complete apeNEXT crate (400 GFlops peak processing power) running at the end of 2002.
Subject to funding availability, we look forward to Tflops class machines
in the year 2003.

\end{document}